\documentclass[lettersize,journal]{IEEEtran}
\usepackage{algorithm}
\usepackage{array}
\usepackage[caption=false,font=normalsize,labelfont=sf,textfont=sf]{subfig}
\usepackage{textcomp}
\usepackage{stfloats}
\usepackage{url}
\usepackage{verbatim}
\usepackage{graphicx}
\usepackage{algpseudocode} 
\usepackage{cite}
\usepackage{amsmath,amssymb,amsfonts}
\usepackage{booktabs,caption}
\usepackage[flushleft]{threeparttable}

\usepackage{placeins}

\usepackage{textcomp}
\usepackage{psfrag}
\usepackage{subfig}
\usepackage{caption}
\usepackage{romannum}

\usepackage{multirow}
\usepackage{textcomp}
\usepackage[table,xcdraw]{xcolor}
\usepackage{tabto}
\usepackage{colortbl}
\usepackage{hhline}
\usepackage{cite}
\usepackage{url}
\usepackage{verbatim}
\usepackage[export]{adjustbox}
\usepackage{pdfpages}
\begin{document}
\title{Quantum ATK Analysis and Detection of Toxic Gases Nitrogen Oxide using Pristine, Defective, and Doped Graphene}
\author{Raju Kumar Yadav,~\IEEEmembership{Student Member,~IEEE,}
 Prince Philip,~\IEEEmembership{Student Member,~IEEE,} 
        Boddepalli SanthiBhushan,~\IEEEmembership{Member,~IEEE}       

\thanks{Raju Kumar and Boddepalli SanthiBhushan are with the Department of Electronics and Communication Engineering
             Indian Institute of Information Technology, Allahabad,  email: rajunptel@gmail.com}
\thanks{Prince Philip is with the Department of Electronic Systems Engineering, Indian Institute of Science, Bangalore, KA, 560012 INDIA (e-mail: princephilip@iisc.ac.in).}%

}
\maketitle
\begin{abstract}
While doping and defects are often considered detrimental to material performance, at the nanoscale, modifications are needed to create novel properties beneficial for device applications. In this work, we focus on optimizing Graphene as a gas sensor for detecting toxic gases such as Nitrogen Oxide (NO). The study explores the effects of doping Graphene sheets with transition metals (Cu, Au, Pt) and introducing a Single Vacancy (SV) defect at the center of the sheet. Pristine, defected, and doped Graphene sheets are systematically analyzed as potential sensing materials for NO gas detection. The investigation includes the design of Graphene-based devices and evaluation of their electrical I-V characteristics under different configurations. The sensing mechanism is examined through parameters such as electronic properties, charge transfer, adsorption energy, electrical characteristics (I-V), sensitivity, and the Non-Equilibrium Green’s Function (NEGF) approach. The results indicate that defected Graphene demonstrates superior gas adsorption performance, with an adsorption energy of 8.316 eV and a sensitivity of 51.1\%, outperforming both pristine and doped Graphene. These findings establish defected Graphene as a promising candidate for NO gas sensing applications, while doped Graphene shows moderate sensing potential.
 
\end{abstract}


\section{Introduction}
Nitric oxide (NO) is a toxic, non-flammable oxidizing gas with a sweet, sharp smell. It is produced through various mechanisms like oxidation with Nitrogenous material, synthesis of nitric acid, and manufacturing in semiconductors\cite{WinNT}. It is harmful to respiratory effects, and environmental hazards. It is essential to detect harmful gases to ensure individual safety and the health of the environment.
Gas sensors specifically designed to detect nitric oxide can provide early warning signs of its presence, allowing appropriate actions to be taken to prevent exposure and mitigate potential health risks\cite{cruz2021recent}.
Gas detection is important for measuring the level of different gases within the air, and is used to prevent anyone from being exposed to toxic gases that could poison, harm, or kill. Chemical sensors have developed considerably over the last few decades and in various fields. Sustainable construction, medical diagnostics, and national defense all heavily rely on gas sensors and gas detection \cite{khan2020sensor}.
Graphene has been the subject of extensive research since its discovery \cite{geim2007rise} because it shows outstanding electrical and thermal properties \cite{mattevi2009evolution} and application in numerous technological fields of sensor \cite{gerasimov2016graphene}, filtration \cite{huang2015graphene}, biomedical application \cite{shen2012biomedical}, wearable electronics, Electronics, and optoelectronics \cite{chen2019graphene}. A single layer of carbon atoms structured in a hexagonal lattice makes up a two-dimensional substance known as graphene \cite{pumera2011graphene}. It is known for having special qualities such as outstanding mechanical and thermal conductivity, flexibility, and electrical conductivity. The significance of graphene comes from its potential to transform a variety of applications and uses \cite{tian2018research}.

Because of its high surface area, high sensitivity, and fast response time, graphene is a usefull material for gas-sensing applications\cite{tian2018research}. The changes in the electrical properties of graphene caused by gas molecules can be detected.
Due to the interaction between the gas molecules and the graphene, changes in the material's electrical conductivity or other electronic characteristics may be the cause of this sensitivity.
A study published in the journal "Nanoscale," for example, demonstrated the use of a graphene oxide sensor to detect NO gas with a detection limit of 50 ppb and a response time of less than 30 seconds \cite{wang2016review}.
For real-time gas sensing applications, where prompt detection and measurement of gas concentrations are necessary, this quick reaction time is crucial.
Overall, graphene is a promising material for gas sensing applications such as NO gas detection. Graphene gas sensors, with further research and development, could have a significant impact on environmental monitoring.

A good number of Graphene-based sensor research work has been done in the research area on the analysis of its density towards Toxic gases. Doped Graphene with gold nanoparticles or a foam-like Graphene model has been experimented with to enhance Ammonia sensing ability \cite{gautam2012ammonia}.
An investigation suggested that Defect Graphene with transition metal increases its sensitivity \cite{katta2022investigation}.
The adsorption of $H_{2}CO$ at the surface of defective Graphene, interactions are present due to the carbon atom of the vacancy area while no interaction in Pristine Graphene \cite{zhou2014dft}. 
This evidence implies that doped and defects improve and enhance the quality of Gas sensors of pristine Graphene.
The gap of Research work is to make an overall analysis of doped and defective Graphene \cite{katta2022investigation,mohan2018graphene}.
A major part of my research paper is based on doped graphene as a gas sensor and defective graphene as a gas sensor. 
Gas sensing has emerged as a crucial area of research, and doped graphene has shown great promise as an effective gas sensor. Among the various types of graphene-based gas sensors, the "doped graphene" gas sensor has been found to exhibit the highest level of performance.

In a subsequent study, defect graphene with different orientations of defect carbon atoms was investigated as an alternative gas sensor configuration. Through comprehensive analysis and comparison, it was determined that defective graphene, with its unique structural imperfections, serves as an exceptional gas sensor. The defective graphene sensor demonstrated remarkable stability under different operating conditions. Consequently, it can be concluded that defective graphene represents an excellent choice for gas sensing applications, offering both high sensitivity and stability. These findings contribute to the advancement of gas sensor technology and provide valuable insights for the development of reliable and efficient sensing devices.
Motivated by these studies, we developed \& designed Graphene with adsorbed NO Gas. Find its all computational and electrical characteristics. Then we got to know about doping improves the sensing behavior of Graphene. Then we doped with different transition elements Copper\/Gold\/Platinum. Then we find after getting the result of the NGF-based Synopsys quantum ATK tool. Silver and platinum-doped graphene show higher sensitivity towards the NO gas.
So we studied/explored vacancy defect Graphene and its sensitivity. So we remove one middle carbon from graphene sheets and adsorbed NO gas on its surface.
As a consequence of vacancy defects, variation in electrical and transport properties has been investigated in large variations. So we finally concluded that vacancy defects show higher sensitivity towards NO Gas.

The rest of the sections are organized as follows. Section \ref{archdigital} describes the methodology. Section \ref{resultsdigital} presents the simulation results and discussion. Finally, the paper is concluded in Section \ref{conclusiondigital}.

\section{Methodology}
\label{archdigital}

 \begin{figure}[t]
	\centering
	\includegraphics[width=1.\linewidth]{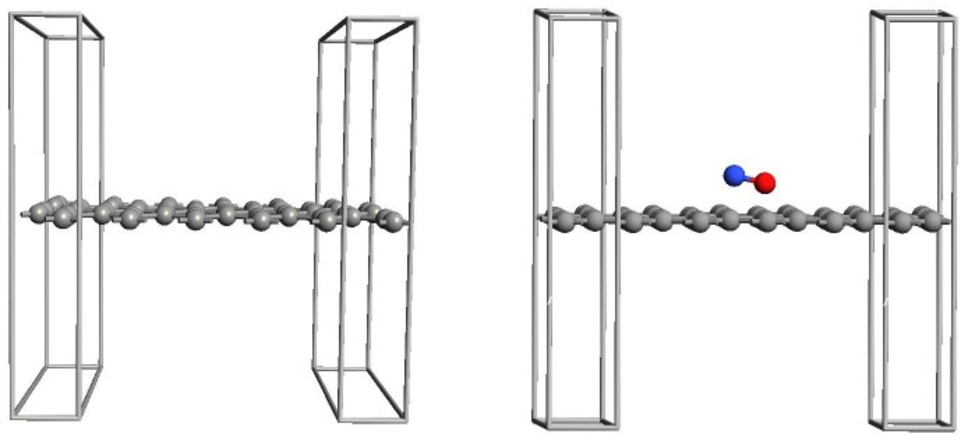}
		\label{retina_model}
  \caption{Graphene device before and after adsorption NO Gas on a Graphene surface.}
	\label{adsorptionNOGas}
\end{figure}
The Present work has been performed using density functional theory and non-equilibrium Green’s function (NEGF)  framework implemented through a quantum ATK -VNL which is a further development of SIESTA \cite{katta2022investigation}. 
To assess the sensing behavior of Graphene sheets towards NO Gas, its geometric structures, electronic and transport properties in pristine, Defected \& doped Graphene sheets are being explored or studied before and after adsorption of NO gas can be used as a sensing material for toxic NO Gas.
The sensing mechanism is studied with the help of its Electronic nature, charge transfer, Density of State, Adsorption Energy, Electrical properties (I-V), Recovery time, and Sensitivity.
The electronic nature helps me to know conducting nature of Graphene sheets with different cases.
Charge transfer can be calculated by quantum ATK to know the mullicane population Adsorption Energy
In graphene research, adsorption energy refers to the energy associated with the binding of an adsorbate (molecule, atom, or ion) to the surface of graphene. It is an important parameter that determines the stability and strength of the interaction between the adsorbate and the graphene surface \cite{dandeliya2018defected}.
For finding the sensing ability to detect the gases, we approached two methods that are computational and electrical I-V characteristic methods.
1. For the modeling of Graphene,32 carbon atom is considered in a single layer in 2-dimensional means a single atomic layer of graphite has been considered. A hexagonal supercell is kept periodic in the x and y directions during optimization. However, confined in the z-direction. This graphene is called pristine graphene. All the configurations are fully relaxed by optimizing the process till 0.01eV/Aforce and 0.01 eV/A stress. The K-Point sampling of (15,15,1) has been considered for accuracy with density cut-off 75 which is the density of the real space grid used for electrostatic calculation.

2. Transport calculations have been performed using NEGF formalism and the quasi-Newton method by considering a two-probe approach. We can use other electrode material electrode, but it is of high contact resistant and lattice mismatch so  A two probe model has been considered that is built, using fixed width graphene sheet extending in Z direction and forming the central region with the right (R) and left electrode(L) electrode region depicted in Fig. \ref{adsorptionNOGas}.

A large K point sampling of (1,1,100) has been considered in the calculation of accomplishing the accuracy in the transport property analysis. We applied five different voltage with the right and left electrode voltage from 0 to 1 voltage of interval 0.25 volt and got I-V sensing behavior.
Our approach to finding sensing behavior by comparing properties is to try first with pristine Graphene before and after NO adsorption and I-V characteristics. Doped and defective make noble material at the microelectronics level. 
So after optimizing the middle carbon SV defect Graphene is optimized and again optimized doped transition metal (Copper, Gold, Platinum) and optimized Single Vacancy (SV)  Carbon defect graphene. So finally go to electronic nature, the density of state,mullicane population, and Transmission spectrum and compare which doping transition metals showing best better adsorption energy.
The Adsorption energy for the pristine and doped transition metals has been calculated using equation(1).
$E_{ad}$= $E_{t}$ (sheet+NO molecule)-$E_{t}$ (sheet)-$E_{t}$ (NO molecule)  -(1)
Here $E_{t}$ (sheet+NO molecule)is the total energy of the relaxed graphene sheet after NO adsorption, while Et (sheet) is the total energy of the optimized graphene sheet and  $E_{t}$ (NO molecule) is the total energy of NO molecule \cite{huang2015graphene}.
Finally, Defect Graphene shows has more adsorption energy so it shows better sensing ability characteristics \cite{dandeliya2018defected}.

\section{Results and Discussion} 
\label{resultsdigital}
To evaluate the sensing behavior towards NO Gas, its geometric structures, and electronic and transport properties in pristine and doped (Cu/Au/Pt) states are studied before and after NO Gas adsorption.
The analysis is discussed in two sections, the first one is computational prediction and the second one is electrical I-V Characteristics.[(14,20)]
Computational prediction indicates the best quality of sensing behavior of Graphene and Electrical I-V characteristic indicates the sensing behavior of Graphene towards harmful gases.
The sensing mechanism is studied with the help of its Electronic nature, charge transfer, Density Of State, Formation Energy, Adsorption energy, Electrical properties (I-V), and Sensitivity.
\begin{figure*}[!t]%
    \centering
    \subfloat[]{\includegraphics[width=0.5\linewidth]{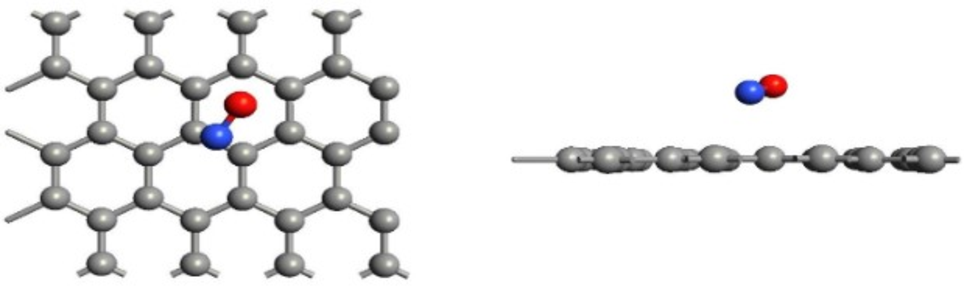}
    \label{GraphenewithNO2}}
     \subfloat[]{\includegraphics[width=0.5\linewidth]{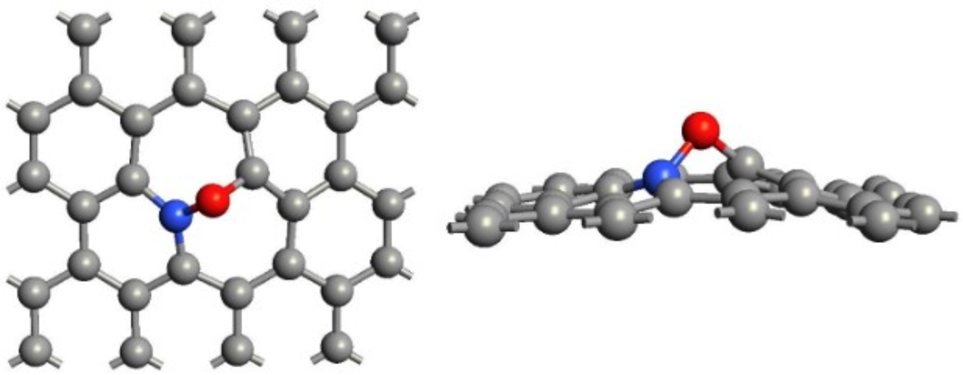}
    \label{defectGraphene2}}\\
  \subfloat[]{\includegraphics[width=0.5\linewidth]{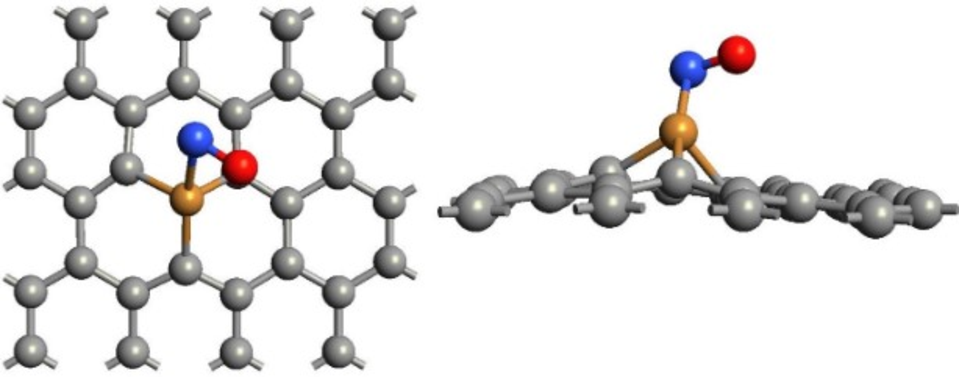}
    \label{DopedGraphene2}}
    \subfloat[]{\includegraphics[width=0.5\linewidth]{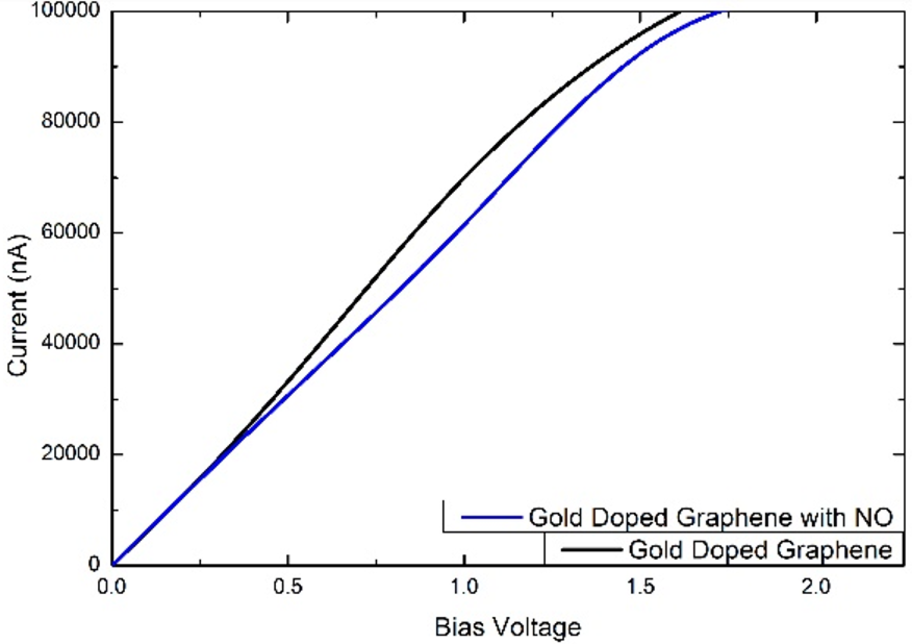}
    \label{GoldDopedGraphene2}}       
       \caption{ \protect\subref{GraphenewithNO2} Top and side view of Graphene with NO; \protect\subref{defectGraphene2} Top and side view of Defective Graphene with adsorbed NO Gas;\protect\subref{DopedGraphene2}Top and side view of Copper Doped Graphene with NO;  \protect\subref{GoldDopedGraphene2} Top view and side view of Gold Doped Graphene with adsorbed NO Gas.}%
    \label{tonicphasic}%
\end{figure*}

\subsection{Pristine Graphene with NO Gas}
The electronic nature of Graphene sheets \cite{gautam2012ammonia} was investigated to determine their conducting properties under various conditions. Initially, the pristine Graphene was found to have a band gap of 0.058 at the quantum ATK tool; however, in reality, its band gap is nearly zero, indicating a quasi-semiconducting behavior. Subsequently, the Graphene sheets were exposed to NO Gas, resulting in metallic. The adsorption of NO Gas introduced localized states that altered the band structure of Graphene.
The Adsorption Energy and sensitivity of Pristine Graphene is 0.77eV and 23 \% which is very low for adsorbing  NO Gas. In Fig. \ref{GraphenewithNO2}, we have attached the Graphene layer with NO gas to analyze the characteristic changes before and after NO gas exposure
\subsection{Defective Graphene with NO Gas}
The Single Vacancy Carbon has been optimized to obtain the relaxed geometry with a C=C bond. In the given fig, the optimized defective Graphene sheet with top and side view with optimized NO gas. The vacancy defect shows a high 8.316 eV (Table \ref{tab2}) adsorption Energy with 51\% sensitivity toward NO Gas, but its formation energy of 0.87 eV (in Table \ref{tabl1}) shows its instability.
\subsection{Copper Doped Graphene with NO Gas}
In this study, we have investigated the impact of copper doping on the electronic properties of Graphene. To achieve copper doping, a single carbon atom in Graphene was replaced to be doped with Graphene with Copper. The design of copper-doped Graphene involved the utilization of optimized Defected Graphene to ensure accurate and reliable analysis. Through extensive electronic structure calculations, we observed a remarkable transformation in the electronic nature of Graphene. Specifically, the introduction of copper doping resulted in a shift from a semiconductor to a metallic behavior. This significant change in the electronic properties of Graphene holds great promise for potential applications in various electronic devices and related fields. The findings presented here contribute to the understanding and exploration of novel materials for advanced electronic applications.
Table \ref{tab2} shows adsorption energy using equation (1) with copper-doped Graphene.
The adsorption energy of Copper Doped Graphene is 2.8027 which is far better than Pristine graphene which improves the sensitivity to NO Gas.
If we explore the sensing ability through sensitivity. It shows the sensitivity of NO gas with Graphene gas. Its sensitivity improves from 23\% to 34\% compared to pristine Graphene.

\begin{equation}
    S\% = \left| \frac{G - G_0}{G_0} \right| \times 100
\end{equation}

Where S (Sensitivity ) $G_{0}$(Conductance before adsorption), G(Conductance after NO adsorption)
In Fig. \ref{DopedGraphene2}, we doped copper in place of a vacant carbon atom to analyze its adsorption capability towards NO gas.
\subsection{Gold Doped Graphene with NO Gas}
Through a comparative analysis between Pristine Graphene and doped Graphene, it was discovered that doping significantly enhances various aspects, including electronic performance \cite{stolbov2015gold}, adsorption energy, and sensitivity towards NO gas. To further enhance the sensing capabilities, a substitution of Gold was made in place of Copper in the Copper Doped Graphene as shown in Fig.\ref{GoldDopedGraphene2}. As a result, the overall sensing ability of the material was greatly improved. The sensitivity towards NO gas increased from 34\% to 49\%, while the adsorption energy rose from 2.8027 eV to 6.565 eV compared from pristine Graphene to Doped Graphene. Moreover, the band gap of the material experienced a transition from metallic nature to 0.22 E/eV after NO adsorption, indicating a shift from metallic to semiconductor behavior. These findings highlight the potential of Copper Doped Graphene as a highly effective sensing material, with improved performance and sensitivity for various sensing applications. The results obtained from this study contribute to the advancement of sensing technologies and provide valuable insights for the development of optimized sensing materials with enhanced properties.
\subsection{Platinum Doped Graphene with NO Gas}
\begin{figure}[t]
	\centering
	\includegraphics[width=1.0\linewidth]{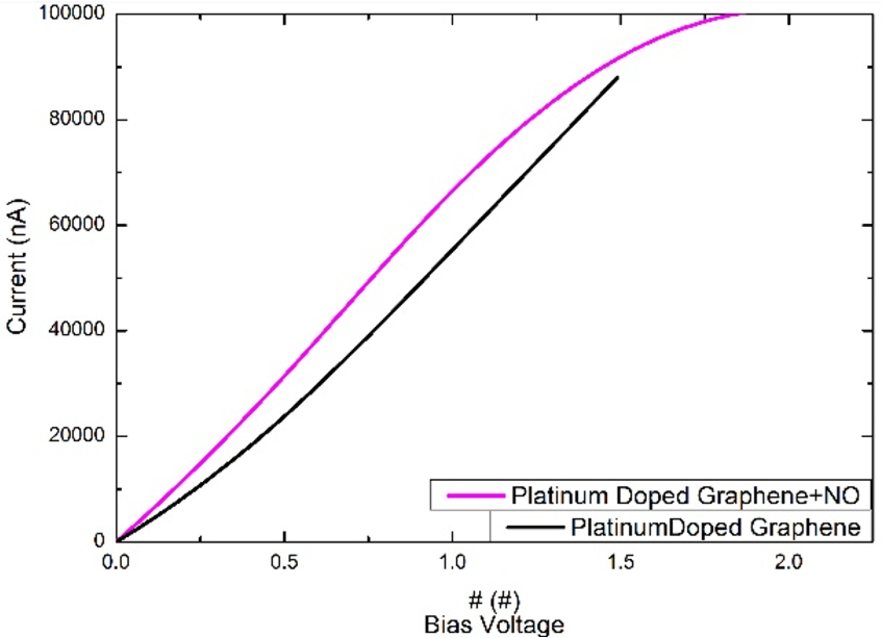}
		\label{GraphenewithNO}
  \caption{\protect Top view and side view of Platinum Doped Graphene with adsorbed NO Gas}
	\label{GraphenewithNO}
\end{figure}
\begin{figure*}[!t]%
    \centering
    \subfloat[]{\includegraphics[width=0.2691\linewidth]{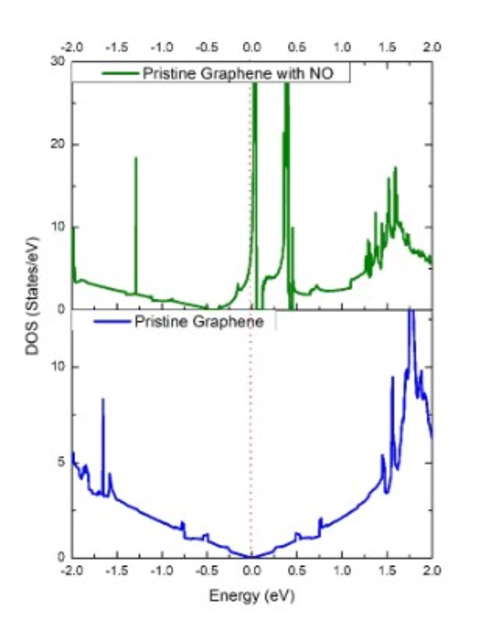}
    \label{pulseinput}}
    \hspace{-0.75cm}
     \subfloat[]{\includegraphics[width=0.2691\linewidth]{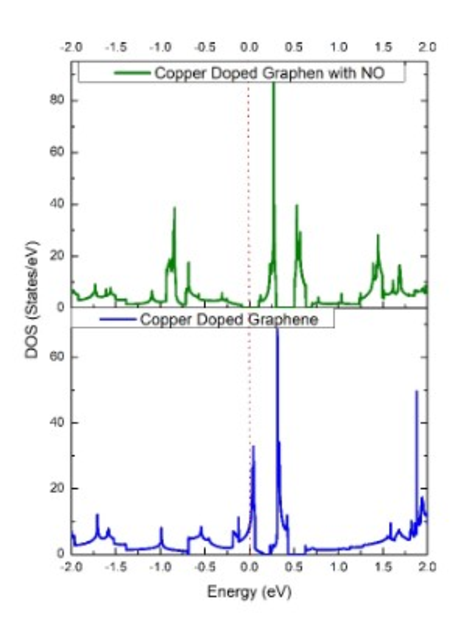}
    \label{singlespike}}
        \hspace{-0.75cm}
  \subfloat[]{\includegraphics[width=0.2691\linewidth]{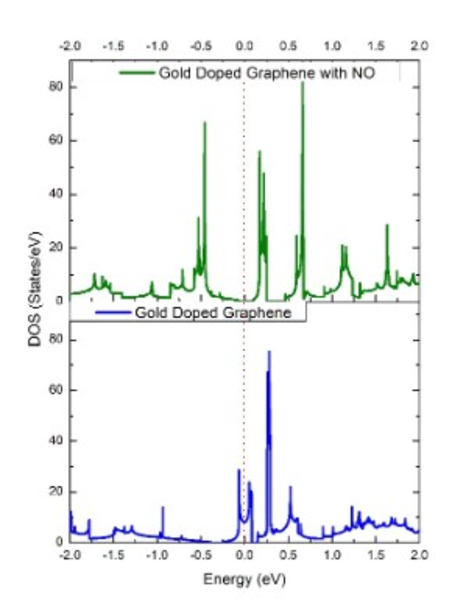}
    \label{tonic}}
        \hspace{-0.75cm}
    \subfloat[]{\includegraphics[width=0.2691\linewidth]{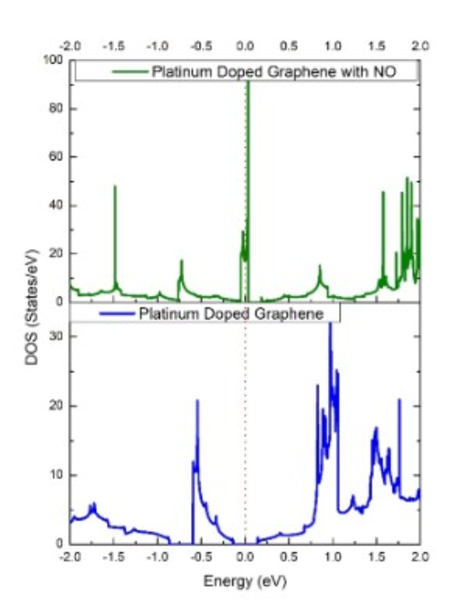}
    \label{phasic}}       
       \caption{ \protect\subref{pulseinput} Density Of State of Pristine Graphene before and after NO gas adsorption; \protect\subref{pulseinput} Density Of State of Copper Doped Graphene before and after NO gas adsorption; \protect\subref{tonic} Density Of State of Gold doped Graphene before and after NO Gas adsorption ; \protect\subref{phasic} Density Of State of Platinum doped Graphene before and after NO Gas Adsorption.}%
    \label{tonicphasic}%
\end{figure*}

\begin{table}[h!]
\centering
\caption{Electronic nature of Graphene before and after NO adsorption, and formation energies before NO adsorption.}
\begin{tabular}{|c|c|c|c|}
\hline
\multirow{2}{*}{Band Gap Type} & \multicolumn{2}{|c|}{Before NO Adsorption (eV)} & \multirow{2}{*}{\parbox{2cm}{Band Gap After \\ NO Adsorption (eV)}} \\ \cline{2-3}
                               & Band Gap (eV)      & {\parbox{1cm}{Formation Energy\\ (eV\/Å)}}      &                          \\ \hline
Pristine Graphene                        &0.06         & -        & Metallic                   \\ \hline

Defected Graphene                        & Metallic        & 0.87         & 0.42                  \\ \hline

Cu-doped Graphene                        &0.15        &0.19        &0.30                  \\ \hline

Au-doped Graphene                      & Metallic        & 0.44         & 0.22                  \\ \hline

Pt-doped Graphene                         & 0.30       & 0.18        & 0.00                \\ \hline
\end{tabular}
\label{tabl1}
\end{table}

In this study, we conducted an extensive investigation into the sensing behavior of platinum (Pt) doped Graphene when exposed to NO gas species. The choice of Pt-doped Graphene was motivated by its unique characteristics and potential for enhancing gas sensing performance. To evaluate its sensing capabilities, the Pt-doped Graphene shows the resulting changes in electrical conductivity were carefully measured and analyzed.
The outcomes of our study revealed that Pt-doped Graphene exhibited exceptional sensitivity \cite{salih2021pt} and selectivity towards the adsorbed NO gas. This can be attributed to the presence of platinum atoms within the Graphene lattice, which facilitated strong adsorption of the gas molecules. Consequently, the adsorption-induced modifications in the electrical conductivity of the material were significant, enabling precise and accurate detection and quantification of the adsorbed gas species.
These findings demonstrate the remarkable potential of Pt-doped Graphene as a highly effective sensing material in terms of formation energy in terms of stability \cite{salih2021computational}. Its ability to detect and differentiate between various adsorbed gas species with great sensitivity and selectivity offers promising prospects for applications in gas sensing technologies. The results obtained from this study contribute to the advancement of gas sensing systems and provide valuable insights for the development of improved sensing materials for diverse industrial, environmental, and healthcare applications.

\textbf{Band structure}
Band structure represents the energy level of solids and it is often used to determine about a material is a conductor, semiconductor, or insulator.
Analyzing graphene's band structure is important for understanding its gas-sensing properties. The energy levels and allowed electronic states of material are represented by bands and bandgaps, and they are commonly referred to as a semiconductor's band structure.
In the context of gas sensing in graphene, the band structure is crucial for understanding how the presence of gas molecules affects the electronic properties of graphene
Density Of State
The number of electronic states per unit energy per unit volume available to electrons in a solid material or device is referred to as the density of states (DOS) in device physics. It is a fundamental idea that is used to explain the electrical characteristics of materials and is essential to comprehend a variety of electronic phenomena.

However the band gap of Graphene is zero, but DFT analysis gives 0.058eV. As a zero bandgap material, graphene has no energy difference between its valence and conduction bands. But when graphene is exposed to some gases, such as nitrogen oxide (NO), the band structure of the material changes, and a bandgap forms. For graphene to be used in electronics, a bandgap must be created since a bandgap is required for the controlling of electron flow \cite{khan2020sensor}. The Fermi level shifts when NO molecules get absorbed on the surface of graphene because they donate electrons to the lattice. The band structure of graphene changes as a result of this change in the Fermi energy level, resulting in the development of a bandgap. The total amount of NO molecules present and the level of electron donation determine the bandgap's size \cite{khan2020sensor}
NO adsorption leads to charge transfer between NO and Graphene which creates the band gap which shifts the position of the fermi level that changes the band structure. Finally, any impurity added into pristine graphene lead to a change in the band structure of Graphene \cite{khan2020sensor}. The Formation energy for the Defected and doped Graphene has been calculated using the following relation.\\
The formation energy of doped graphene ($E_{f}$)\\
\begin{equation}
= \frac{E_{Doped Graphen}-N*E_{c}-E_{dopant}}{L_{Length of Graphen sheet}}
\end{equation}
The formation energy of defective  graphene ($E_{f}$)\\
\begin{equation}
= \frac{E_{defective Graphen}-N*E_{c}}{L_{Length of Graphen sheet}}
\end{equation}

\begin{table}[]
\centering
\caption{Calculation of Adsorption Energy and Charge Transfer}
\begin{tabular}{ |p{1.75cm}|p{1.75cm}|p{1.75cm}|p{1.75cm}|   }
\hline
\hline
Optimized Geometry& |$E_{ad}$|(eV) Adsorption Energy &Q(e)Charge Transfer&Charge Transfer from\\
\hline
\hline
Pristine Graphene & 0.770&0.066&NO to Graphene   \\ \hline
Cu-Doped
Graphene 
 & 2.802&0.067&NO to Graphene   \\ \hline
Au-doped Graphene
 & 6.656 &0.091&Graphene to NO   \\ \hline
 Au-doped Graphene & 3.201&0.031&Graphene to NO \\ \hline
 Defective Graphene  & 8.316&0.032&Grapehen to NO \\ 
\hline
\end{tabular}
 \label{tab2}
\end{table}

Where $E_{c}$ is the total energy of single carbon,$E_{(Doped Graphen)}$ is the total energy of a doped Graphene sheet, $E_{(Defected Graphen)}$ is the total energy of defective Graphene, and N is the total no of carbon atoms, $E_{dopant}$ is the total energy of dopant atom.
The stability of a material is strongly influenced by its formation energy, which serves as an important determinant. In general, the stability of a material is inversely proportional to its formation energy. In the case of Defected Graphene, it exhibits a relatively high formation energy of 0.868 eV/Å. To explore alternative materials with enhanced stability, our investigation turned toward doped Graphene, which possesses lower formation energy. By reducing the formation energy, we aim to achieve improved stability and further explore its potential sensing behavior. The investigation into the formation energies of different materials provides valuable insights for the design and development of stable and efficient materials for various applications.

\textbf{Adsorption Energy}:
The Adsorption Energy between molecules and graphene sheet has been calculated using the following expression.
\begin{equation}
\centering
\begin{aligned}
    E_{ad}=E_{t}(sheet+NO molecule)-E_{t}(sheet)\\-E_{t}(NO molecule)
\end{aligned}
\end{equation}

\begin{figure}[t]
	\centering
	\includegraphics[width=1.0\linewidth]{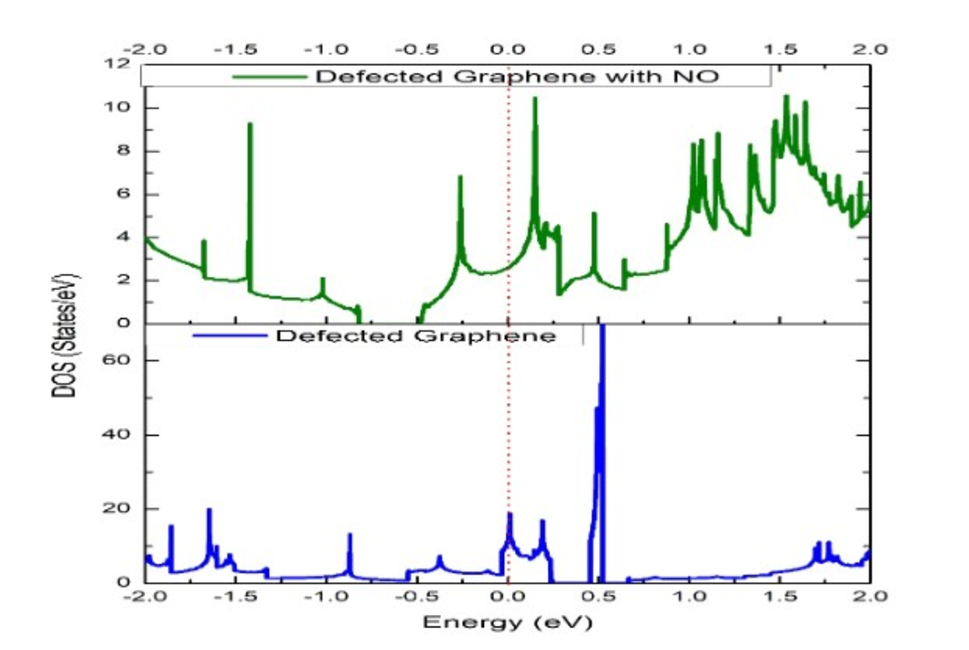}
		\label{GraphenewithNO}
  \caption{\protect Density Of State of Defected Graphene before and after NO adsorption on a Graphene surface}
	\label{GraphenewithNO}
\end{figure}

\begin{figure}[t]
	\centering
	\includegraphics[width=1.0\linewidth]{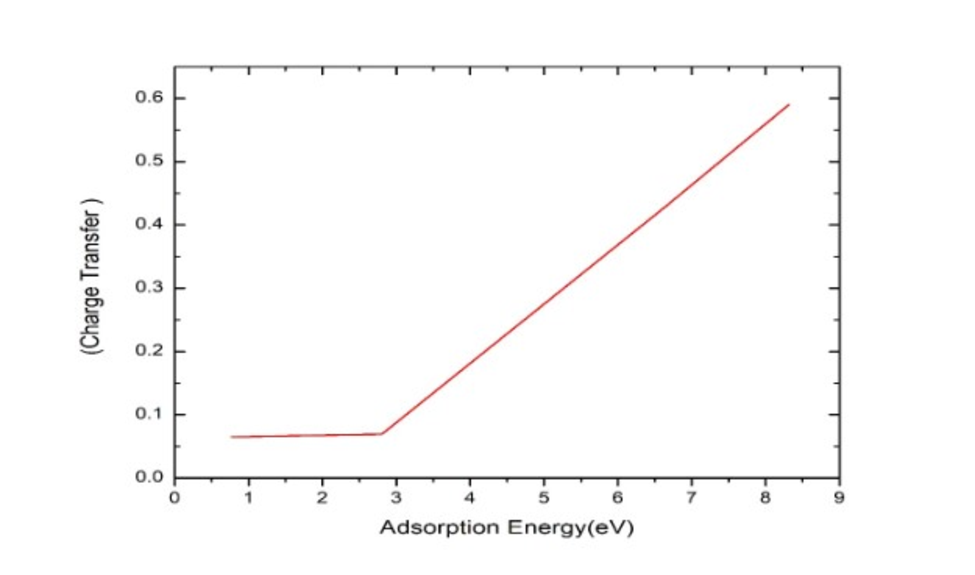}
		\label{GraphenewithNO}
  \caption{\protect Charge Transfer VS Adsorption Energy}
	\label{GraphenewithNO}
\end{figure}

 \begin{figure*}[]
	\centering 
	\subfloat[]{\includegraphics[width=0.34\linewidth]{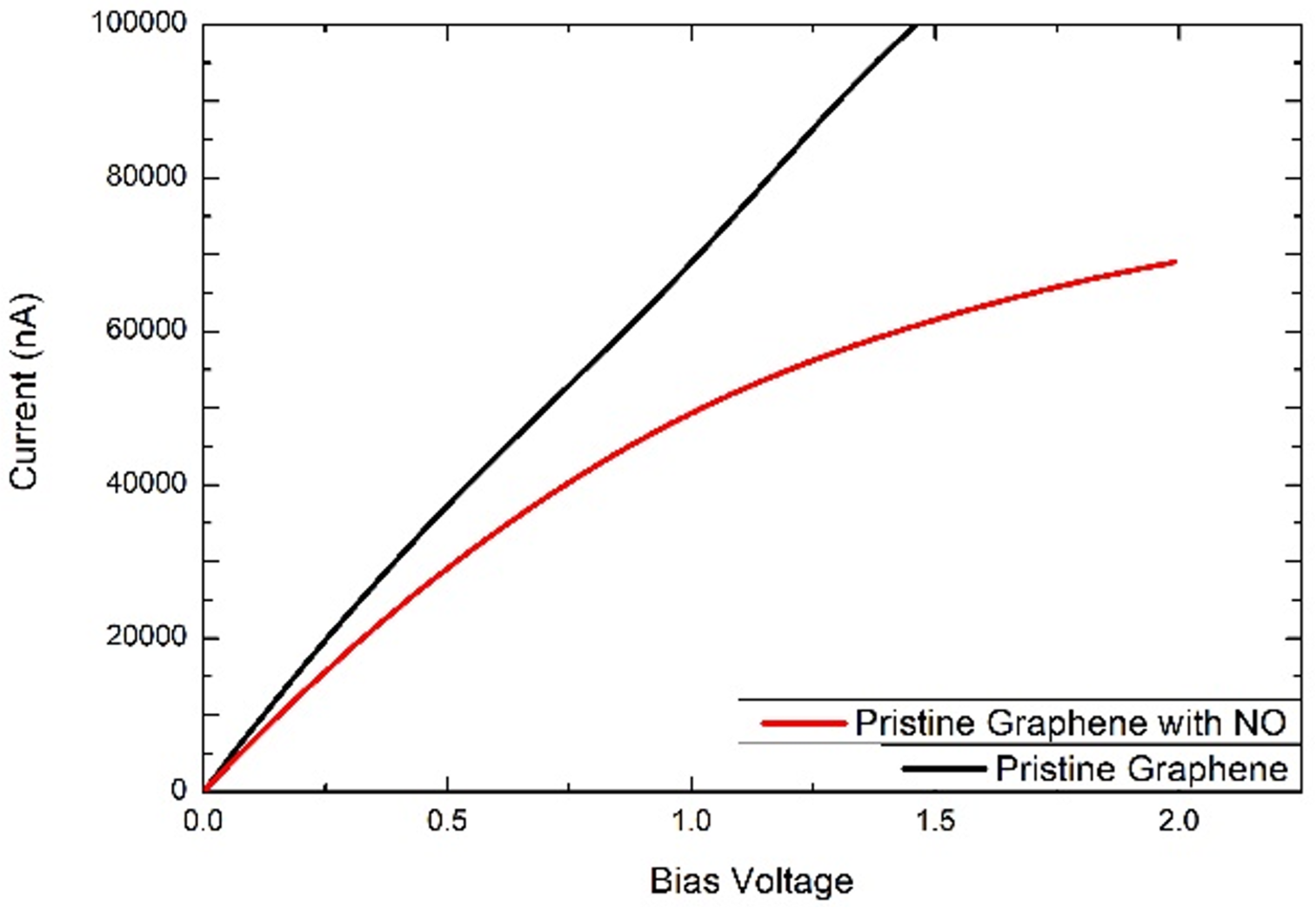}
		\label{PristineGraphene}}
  \subfloat[]{\includegraphics[width=0.34\linewidth]{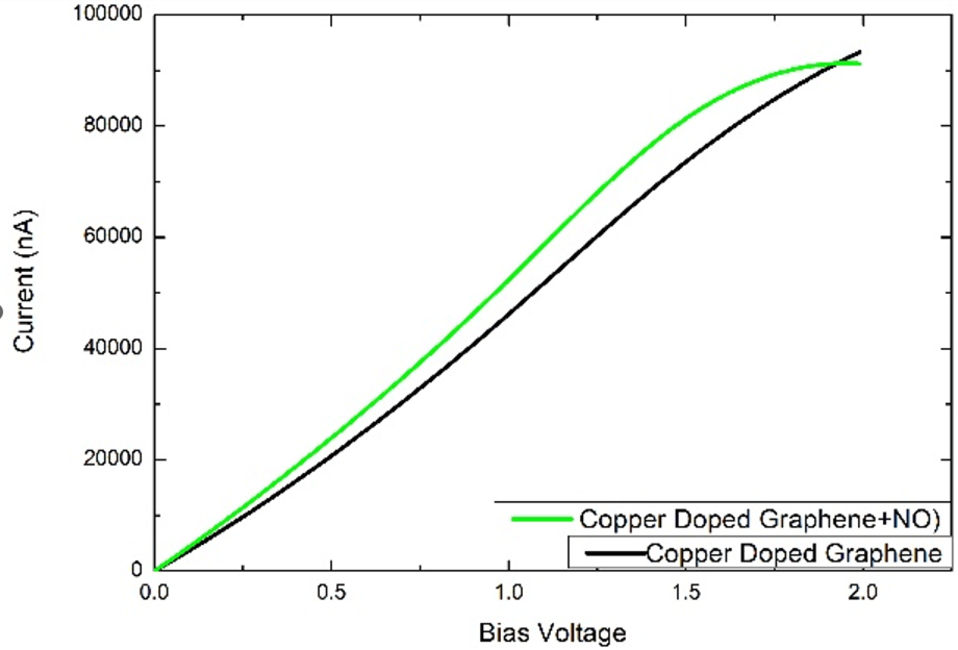}
		\label{CopperDopedGraphene}}
	\subfloat[]{\includegraphics[width=0.33\linewidth]{images/GoldDopedGraphene.eps}
		\label{GoldDopedGraphene}} \\ 
  \subfloat[]{\includegraphics[width=0.347\linewidth]{images/PlatinumDopedGraphene.eps}
		\label{PlatinumDopedGraphene}}
	\subfloat[]{\includegraphics[width=0.315\linewidth]{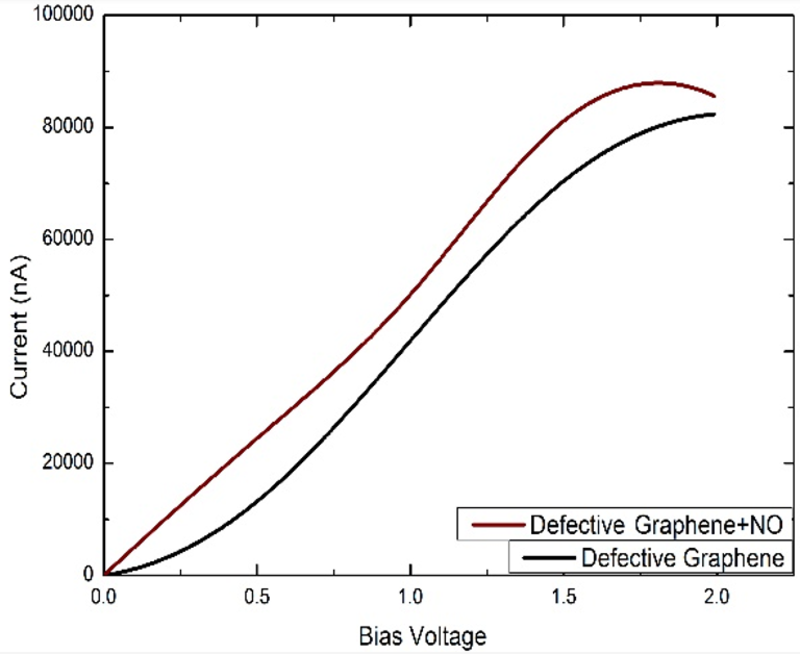}
		\label{DefectiveGraphene}}		
  \subfloat[]{{\includegraphics[width=0.34\linewidth]{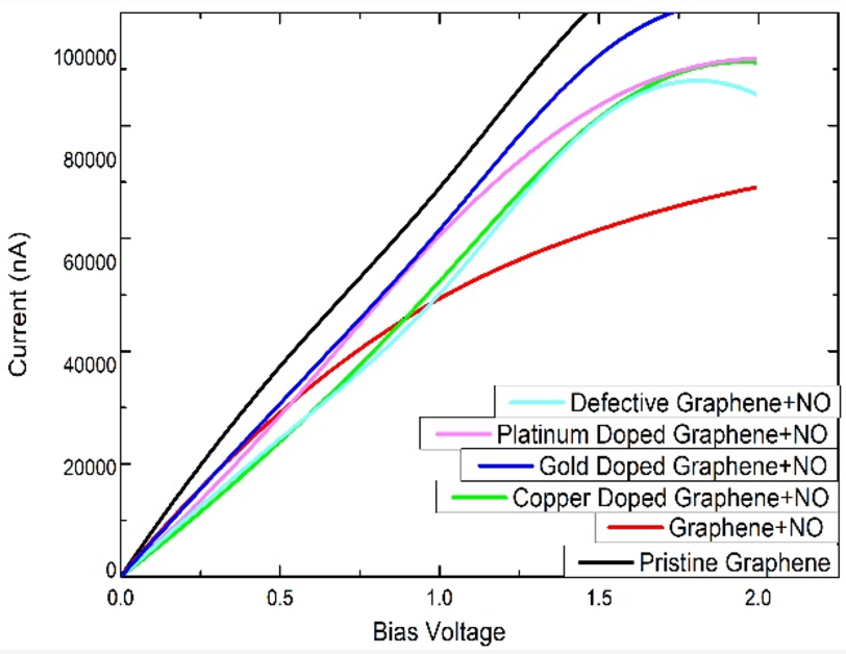}}
		\label{pristineDopedCu}}
 \caption{\protect\subref{PristineGraphene} I-V characteristics of Pristine Graphene;
\protect\subref{CopperDopedGraphene} I-V characteristics of Copper Doped Graphene;
\protect\subref{GoldDopedGraphene} I-V characteristics of Gold Doped Graphene before and after NO adsorption;
\protect\subref{PlatinumDopedGraphene}I-V characteristics of Platinum Doped Graphene before and after NO adsorption
\protect\subref{DefectiveGraphene} I-V characteristics of Defective Graphene before and after NO adsorption;
\protect\subref{pristineDopedCu} I-V characteristics of pristine, Doped(Cu,Au,Pt) \& Defective Graphene before after NO adsorption;
}

	\label{imageout}
 
\end{figure*}

Here $E_{t}$ (Sheet+NO molecule) is the total energy of the relaxed graphene sheet after NO adsorption, while Et (sheet) is the total energy of the optimized graphene sheet and Et (NO molecule) is the total energy of NO molecule \cite{katta2022investigation}. Adsorption energy plays an important role in understanding the interaction strength between an adsorbate, such as a molecule or atom, and the adsorbent surface. It quantifies the energy required to adsorb the species onto the surface. The adsorbate refers to the molecule or atom being adsorbed, while the adsorbent represents the surface onto which the species is adsorbed. The adsorption energy serves as a measure of the binding strength and stability of the adsorbate-adsorbent system. Higher adsorption energies indicate stronger interactions and tighter binding, implying a more stable adsorption process. Studying the adsorption energy is essential for various fields, including surface science, catalysis, and materials science, as it provides valuable insights into the adsorption behavior and can guide the design of efficient adsorbent materials for various applications.
\begin{table}[]
\centering
\caption{Conductance before and after NO adsorption at zero bias voltage and its sensitivity}
\begin{tabular}{ |p{1.75cm}|p{1.75cm}|p{1.75cm}|p{1.75cm}|   }
\hline
\hline
Optimized Geometry& $G_{0}$(Conductance before adsorption) &$G$(Conductance after NO adsorption)&S \% (Sensitivity) \\
\hline
\hline
Pristine Graphene & 7.747238 &5.96387&23   \\ \hline
Cu Doped
Graphene 
 & 3.2039577 &4.320266&34  \\ \hline
Gold Doped
Graphene
 & 3.2334653 &3.392995&49  \\ \hline
 Platinum Doped Graphene & 3.2528756&4.87727920&49  \\ \hline
 Defective Graphene  & 6.073844&9.1782356&5111\\
\hline
\end{tabular}
 \label{tab1}
\end{table}

The Charge transfer from NO molecule is calculated using Mullicane population analyzed by analyzing the difference in charge concentration before and after the NO molecule adsorption.

The energy needed to adsorb an atom or molecule onto a surface is known as adsorption energy.
When a molecule or atom adsorbs onto the surface of graphene, it can interact with the electrons in the graphene lattice, resulting in a transfer of charge between the adsorbate and the graphene surface. This charge transfer can affect the adsorption energy of the system. As much as the adsorbate is nearer to the adsorbent, more charge transfer takes place. So its adsorption energy will be directly proportional to charge transfer.[23]
Some studies have reported that when the same nature, size, same orientation, and electronics nature Silver and Platinum-doped graphene materials exhibit higher reactivity of adsorbed gas towards the graphene sheet compared to Copper and Gold-doped graphene. This can be attributed to the electronic and catalytic properties of Silver and Platinum, which can enhance the gas-sensing performance of the graphene material \cite{hadsadee2022theoretical}.

The degree of the interaction between an adsorbate and the graphene surface normally determines the adsorption energy of an adsorbate on the material. Larger adsorption energy leads to a stronger interaction. The electronic characteristics of the adsorbate and the graphene surface, as well as the separation between them, determine the degree of the interaction.[23]
In conclusion, charge transfer between the adsorbate and the graphene surface can affect the intensity of the interaction between the adsorbate and the graphene surface, which in turn affects the adsorption energy of an adsorbate on graphene [23]. The sensitivity of Graphene can be analyzed by measuring the change in electrical conductivity of graphene before and after the adsorption of NO gas at zero biased voltage. Sensitivity shows how much our adsorbate is sensible to the adsorbent. We have taken different cases to study which cases will be best suited to the sensor.

The Pt-doped Graphene shows higher sensitivity but defect graphene shows the highest sensitivity. My study shows that at the doping level, sensitivity can vary depending on the type of dopant used. It has been demonstrated that defective graphene is more sensitive to NO gas than doped graphene. This is due to the prospect that flaws in the graphene lattice may produce local states in the bandgap that interact strongly with NO molecules, changing the electrical conductivity of the material \cite{tiwari2020graphene}.
Overall, it has been established that defective graphene exhibits better sensitivity than doped graphene for the detection of NO gas because it contains localized states in the bandgap that may interact more strongly with the gas molecules

\subsection{ Electrical Characteristics}
In the field of sensor technology, the evaluation of I-V characteristics provides valuable insights into the response time of Graphene-based sensors. The term "response time" refers to the duration required for a sensor to detect and quantify changes in analyte concentration or other relevant physical parameters \cite{barhoum2023fundamentals}. By analyzing the transient behavior of the I-V curve during the sensing process, it is possible to estimate the response time of the sensor. This information is crucial for assessing the speed and efficiency of Graphene sensors in real-time applications. The I-V characteristics serve as a useful tool in quantifying the response time, enabling researchers to optimize sensor design and enhance sensing capabilities.
To understand the sensing ability, it is important to analyze computational parameters. The sensing mechanism is studied based on its electronic nature, charge transfer, density of states (DOS), adsorption energy, electrical properties (I-V characteristics), and sensitivity. The electronic nature helps in determining the conducting behavior of Graphene sheets in different cases, whereas the I-V characteristics illustrate the change in current with varying biasing voltage.
In Fig. \ref{PristineGraphene}-\ref{DefectiveGraphene}, we observe changes in the electrical properties before and after doping with NO gas.

In this study, we analyzed the sensing behavior of doped graphene within a specific voltage range of 0.5 volts to 1.8 volts, as depicted in Fig. \ref{pristineDopedCu}. Our investigation revealed a distinct pattern in which the current exhibited a rapid and pronounced change in response to the applied voltage. Notably, at a biased voltage of 1.6 volts, a significant variation in the current was observed. This observation emphasizes the potential of doped graphene as a highly responsive material for sensing applications. The outcomes of our research contribute to the advancement of graphene-based sensors, enabling the development of novel and efficient sensing devices with enhanced performance and reliability.
\section{Conclusion}
\label{conclusiondigital}
This work analyzed pristine graphene, transition-metal-doped graphene (Copper, Gold, Platinum), and defective graphene with single-vacancy configurations for their NO gas sensing capabilities. While pristine graphene exhibited limited reactivity, transition-metal doping significantly improved its sensing performance. Among the doped variants, Pt-doped graphene demonstrated the highest stability due to its favorable formation energy. However, defective graphene with single-vacancy defects emerged as the most effective sensor, offering superior sensitivity and enhanced charge transfer properties. The introduction of defects not only increased adsorption energy but also facilitated additional charge transfer pathways, resulting in exceptional sensing performance.
Furthermore, gold-doped graphene demonstrated better sensing characteristics compared to pristine and other doped graphene. Overall, defective graphene proved to be the most promising candidate, combining high sensitivity and efficient charge transfer. These insights advance the understanding of graphene-based sensing materials and contribute to the development of high-performance sensors for diverse applications.

\section*{Acknowledgment}
The author would like to express their graduate to the Indian Institute of Information Technology Allahabad for providing support to write the paper to carry out the present research work. A special thanks to Boddepalli santhiBhushan and Sanjay Singh ECE Department for the inspiration and support and for providing the necessary infrastructure to carry out current research.
\section*{Conflict of Interest Statement}
The authors have no conflicts of interest to disclose.

\FloatBarrier
\bibliographystyle{elsarticle-num}
\bibliography{ref}
\FloatBarrier

\vfill

\end{document}